\documentstyle[12pt]{article}
\begin{document}
\renewcommand{\thefootnote}{\fnsymbol{footnote}}
\thispagestyle{empty}
\begin{center}
{\large\bf
Partial supersymmetry breaking in Multidimensional N=4 SUSY QM.
}\vspace{0.2cm} \\

E. E. Donets,\footnote{E-mail: edonets@sunhe.jinr.ru}
\vspace{0.2cm} \\
{\it  JINR -- Laboratory of High Energies,
 141980 Dubna, Moscow region, Russia, }  \vspace{0.2cm} \\

A. Pashnev,\footnote{E-mail: pashnev@thsun1.jinr.dubna.su}
J.J.Rosales,\footnote{E-mail: rosales@thsun1.jinr.dubna.su}
and  M. Tsulaia\footnote{E-mail: tsulaia@thsun1.jinr.dubna.su}
\vspace{0.5cm} \\
{\it JINR--Bogoliubov Theoretical Laboratory,         \\
141980 Dubna, Moscow Region, Russia} \vspace{1.5cm} \\
\end{center}
\vspace{.2cm}

\begin{abstract}
The multidimensional $N=4$ supersymmetric quantum mechanics (SUSY QM)
is constructed 
and the various possibilities
for partial supersymmetry breaking are discussed. 
It is shown that  quantum
mechanical models with one quarter, one half and three quarters of
 unbroken(broken) supersymmetries can exist in the framework of the
multidimensional $N=4$ SUSY QM. \\
\end{abstract}
\begin{center}
{\it Talk given at the International Workshop SQS99 in Dubna.}
\end{center}
\newpage
\renewcommand{\thefootnote}{\arabic{footnote}}
\setcounter{footnote}0
The supersymmetric quantum mechanics (SUSY QM), being first introduced
in  \cite{WW} -- \cite{ED} for the $N=2$ case, turns out to be a
convenient
tool for investigating problems of  supersymmetric field theories,
since it provides the simple and, at the same time, quite adequate
understanding of various phenomena arising in  relativistic theories.

The important question of all modern theories of fundamental interactions,
including superstrings and M~ -- theory,
is the problem of  spontaneous breaking of supersymmetry.
Supersymmetry, as a fundamental symmetry of the nature, if exists,
has to be spontaneously broken at low energies since particles with
all equal quantum numbers, except the spin, are not observed
experimentally.

The problem of  spontaneous breaking of supersymmetry
could be investigated in the framework of the supersymmetric quantum
mechanics as well. 
The one -- dimensional $N=4$ SUSY QM was constructed first in
\cite{BP1} -- \cite{IKP}.  Partial breaking of supersymmetry, caused by
the presence of the
central charges in the corresponding superalgebra, was
also discussed in  \cite{IKP}. It was the first example of  partial
breaking of supersymmetry in the framework of SUSY QM and the corresponding
mechanism is in full analogy with that in  \cite{ANTONIAD}  -- \cite{IZ}
 in the field theory. The main point is that the presence
of  central charges in the superalgebra allows
the partial supersymmetry breaking, whereas according to
Witten's theorem \cite{WW}, no partial supersymmetry breaking is
possible if the SUSY algebra includes no central charges.
The main goal of our paper is further generalization of the construction,
proposed in  \cite{IKP} to the multidimensional case and
investigation of
partial breaking of supersymmetry under  consideration.

 Consideration of the supersymmetric algebra with
central charges is of particular importance  for  several reasons.
 First, it provides a good tool to
study  dyon solutions of quantum field theory since in such theories
the mass and electric and magnetic charges  turns out
to be the central charges \cite{WO}. Second,  the central
charges produce the rich structure of supersymmetry breaking. Namely,
it is possible to break  part of all supersymmetries retaining all others
exact \cite{WESS}.
In fact, the invariance of a state with respect to the supersymmetry
transformation means  saturation of the Bogomol'ny bound, and this
situation takes place in $N=2$ and $N=4$ supersymmetric Yang -- Mills
theory \cite{GAU} as well as in the theories of  extended
supergravity \cite{KLO}.

 The investigation of
supersymmetric properties of branes in the  M~ -- theory
has also revealed that
 partial breaking of supersymmetry takes place. Namely, the ordinary
branes break half of the supersymmetries, while ``intersecting" and
rotating
branes can leave only $1/4$,  $1/8$, $1/16$ or $1/32$ of the
supersymmetries unbroken
\cite{STT}.

The main characteristic features of   partial SUSY breaking
in the field theories with the extended supersymmetry can be revealed
in supersymmetric QM, since in the both cases  partial supersymmetry
breakdown is provided by the central charges in the SUSY algebra.
Therefore the detailed study of  partial supersymmetry
breakdown in supersymmetric quantum mechanics can lead to the deeper
understanding of analogous effect in supersymmetric field theories.

Let us describe a general formalism of
the $D$~-- dimensional
($D \ge 1$) $N=4$ supersymmetric
quantum mechanics \cite{DP}.
The physical content of the theory is: 
the bosonic fields $\phi^i$ and fermionic fields $\psi^{ai}$ and
$\bar \psi_a^i$, where $i = 1,...,D$ and index $a=1,2$ is $SU(2)$
group index.\footnote{Our conventions for spinors are
as follows:
${\psi}_{a}
={\psi}^{b}{\varepsilon}_{ba},\; {\psi}^{a}=
{\varepsilon}^{ab}{\theta}_{b},\;
{\bar \psi}_{a}={\bar \psi}^{b}{\varepsilon}_{ba},\; {\bar \psi}^{a}=
{\varepsilon}^{ab}{\bar \psi}_{b},\;
{\bar \psi}_a = (\psi^a)^\ast,\;
{\bar \psi}^a = -(\psi_a)^\ast,\;
\varepsilon^{12} = 1,\;
\varepsilon_{12} = 1$.}
The hamiltonian and the supercharges, desribing the multidimensional
N=4 SUSY QM have the form:

\begin{equation}  \label{QUANT1}
\bar Q_a=\bar \psi^i_a R_i  - 2i \bar \psi^i_a m^j
(\partial^2_{ij}  A)
+2i \psi_a^i n^j (\partial^2_{ij} A)
- \frac{1}{2}i \lambda^c_{ai} \bar \psi^i_c ,
\end{equation}
\begin{equation}
Q^b=L_i\psi^{bi}  + 2i \psi^{bi} m^j \label{QUANT2}
(\partial^2_{ij}   A)
+2i \bar \psi^{bi} \bar n^j (\partial^2_{ij}  A)
+\frac{1}{2}i \lambda^b_{di} \psi^{di} ,
\end{equation}
\begin{eqnarray}  \nonumber
H_{quant.}&=&\frac{1}{2}L_i
{(\partial^2_{ij} A)}^{-1}R_j
+ \frac{1}{16}\lambda^a_{bi}\lambda^{b}_{aj}
{(\partial^2_{ij} A)}^{-1}  +
2 (\partial^2_{ij} A)
(m^im^j +  n^i\bar n^j) + \\ \nonumber
&+& (\partial^3_{ijk} A)
( [\bar \psi^i_a \psi^{aj}] m^k + \psi^{ai} \psi_{a}^{j} n^k +
\bar \psi^{i}_{a} \bar \psi^{aj} \bar n^k) - \\ \nonumber
&-&\frac{1}{4} \lambda^a_{bp}
 {(\partial^2_{pk} A)}^{-1}
(\partial^3_{ijk} A)
[\bar \psi^i_a , \psi^{bj}]
+\frac{1}{2}
(\partial^4_{ijkl} A)
(\bar \psi^i_a \bar \psi^{ja})(\psi^{bk} \psi_b^l) -  \\ 
\label{OPERATOR}
&-& {(\partial^2_{pk} A)}^{-1}
((\partial^3_{ikp} A)
(\partial^3_{qjl} A)
+ (\partial^3_{ijp} A) (\partial^3_{qkl} A))
\bar \psi^j_{a} \bar \psi_b^k \psi^{bl} \psi^{ai} ,
\end{eqnarray}
where $m^i$ is a set of real constants, $n^i$ and $\bar n^i$ are 
mutually complex conjugated constants, $\lambda^a_{bi}$ is a set
of $SU(2)$ -- valued constant matrices,
$A(\phi^i)$ is an arbitrary function, called the superpotential and
\begin{eqnarray} \nonumber
L_i&=&p_i + i \bar \psi_a^j \psi^{ak}
(\partial^3_{ijk} A)
-\frac{i}{2}
{(\partial^2_{jk} A)}^{-1}
(\partial^3_{ijk} A) , \\
R_i&=&p_i - i \bar \psi_a^j \psi^{ak}
(\partial^3_{ijk} A)
+\frac{i}{2}
{(\partial^2_{jk} A)}^{-1}
(\partial^3_{ijk} A) .
\end{eqnarray}

~The ~momentum ~operators ~are ~Hermitean ~with ~respect ~to
~the ~integration ~measure
$d^D \phi \sqrt{| det(\partial^2_{ij} A)|}$
if they have the following
form:
\begin{equation} \label{IMP}
p_i = -i \frac{\partial}{\partial \phi^i} -
\frac{i}{4} \frac{\partial}{\partial \phi^i} 
ln (| det (\partial^2_{ik}) A|)
-2i \omega_{i \alpha \beta}\bar \psi^{\alpha}_a \psi^{a \beta} ,
\end{equation}
with the new fermionic variables $\bar \psi^{\alpha}_a$  and
$ \psi^{a \beta}$ connected with the old ones via the tetrad
$e^\alpha_i$ ($e^\alpha_i e^\beta_j \eta_{\alpha \beta} = 
\partial_{ij}^2A$)
\begin{equation} \label{TETRAD}
\bar \psi^{\alpha}_a = e^\alpha_i \bar \psi^{i}_a \quad
\hbox{and} \quad
 \psi^{\alpha}_a = e^\alpha_i  \psi^{i}_a ,
\end{equation}
and  $\omega_{i \alpha \beta}$ in (\ref{IMP}) is the spin
connection, which corresponds to the metric
$\partial^2_{ij} A$.
These operators form the following N = 4 SUSY algebra with 
the central charges
$$
\{ \bar Q_a , Q^b \} =  \delta^b_a H_{quant.} +   \lambda^b_{ai}m^i ,
$$
\begin{equation} \label{OOO}
\{ \bar Q_a , \bar Q_b \} =   \lambda_{abi}n^i , \quad
\{  Q^a , Q^b \} =  -  \lambda^{ab}_i \bar n^i ,
\end{equation}
with respect to the  (anti)commutators
$$
[ \phi^i, p_j ]  = i \delta^i_j ,
\quad
\{ \psi^{ai}, \bar \psi^j_b \} =  \frac{1}{2} \delta^a_b
{(\partial^2_{ij} A)}^{-1} 
$$
$$
[ \psi^{ai}, p_j ]=- \frac{i}{2}  \psi^{ap} (\partial^3_{pmj} A)
{(\partial^2_{mi} A)}^{-1} , \quad
[ \bar \psi^i_a, p_j ] = - \frac{i}{2}  \bar \psi^p_a
(\partial^3_{pmj} A)
{(\partial^2_{mi} A)}^{-1} 
$$

and
$$
[ p_i, p_j ]= \frac{1}{2}
{(\partial^2_{pq} A)}^{-1}
((\partial^3_{ikp} A)(\partial^3_{qjl} A) -
(\partial^3_{ilp} A)(\partial^3_{qjk} A))
\bar \psi_a^k \psi^{al} 
$$

Let us investigate in detail the question of  partial supersymmetry
breaking in the framework of the constructed $N=4$ SUSY QM in
an arbitrary number of dimensions $D$. 
We shall see that in contrast with the one -- dimensional $N=4$
SUSY QM, the multidimensional one provides also
 possibilities when either only one
quarter of all supersymmetries is exact (for $D \geq 2$),
or one quarter
of all supersymmetries is broken (for $D \geq 3$).

In order to study partial SUSY breaking it is convenient
to introduce a new set of real -- valued supercharges
$$
S^a = \bar Q_a +  Q^a ,
$$
$$
T^a = i(\bar Q_a -  Q^a) .
$$
and the label ``a" has now to be
considered just as the number of  supercharges denoted by $S$ and $T$.

The new supercharges form the following $N=4$ superalgebra with
the central charges
\begin{eqnarray} \label{ALGEBRA1}
\{ S^a, S^b \}&=& H (\delta ^a_b +\delta^b_a)
+  (\lambda^a_{bi} +\lambda^b_{ai})m^i +
(\lambda _{abi} n^i - \lambda^{ab}_i \bar n^i) ,  \\  \label{ALGEBRA2}
\{ T^a, T^b \}&=& H (\delta ^a_b +\delta^b_a)
+  (\lambda^a_{bi} +\lambda^b_{ai})m^i -
(\lambda _{abi} n^i - \lambda^{ab}_i \bar n^i) , \\  \label{ALGEBRA3}
\{ S^a, T^b \}&=&i (\lambda^a_{bi}  - \lambda^b_{ai})m^i +
i(\lambda _{abi} n^i + \lambda^{ab}_i \bar n^i) ,
\end{eqnarray}
where $\lambda^{a}_{bi}=(\sigma_I)^a_b \Lambda^I_i$ and
$\Lambda^I_i$ are real parameters.

The algebra (\ref{ALGEBRA1}) -- (\ref{ALGEBRA3}) is still
nondiagonal. However, some particular choices of the
constant parameters $m^i, n^i$ and $\Lambda_{i}^I$ bring
the algebra to the standard form, i.e., to the form
when the right -- hand side of (\ref{ALGEBRA3}) vanishes and
the right -- hand sides of (\ref{ALGEBRA1}) and (\ref{ALGEBRA2})
are diagonal with respect to the indices ``a" and ``b".

Now we consider several cases separately.

{\bf Four supersymmetries exact / Four supersymmetries broken.}
If we put equal to zero all central charges, appearing
in the algebra, then no partial breaking of supersymmetry
is possible. In this case, all supersymmetries are exact,
if the energy of the ground state is zero; otherwise all of them are
broken. This statement is obviously independent of the number of
dimensions $D$.

{\bf Two supersymmetries exact.}
The case of  partial supersymmetry breaking,
 when the half of  supersymmetries are exact, have been
considered earlier \cite{IKP} in the framework of one-dimensional
$N=4$ SUSY QM,  but we shall describe it  for completeness as well.
Consider the one -- dimensional ($D=1$) $N=4$ SUSY QM and put all
the constants entering into the right -- hand sides
of (\ref{ALGEBRA1}) -- (\ref{ALGEBRA3})
equal to zero, except $m^1$ and $\Lambda^3_1$.
Then, the algebra (\ref{ALGEBRA1}) -- (\ref{ALGEBRA3}) takes the form
\begin{eqnarray}   \nonumber
\{ S^1 , S^1 \}&=&\{ T^1 , T^1 \}=2H + 2 m^1  \Lambda^3_1 , \\
\{ S^2 , S^2 \}&=&\{ T^2 , T^2 \}=2H - 2 m^1  \Lambda^3_1 ,\label{TWO}
\end{eqnarray}
It means that if the energy of the ground state is equal
to $m^1  \Lambda^3_1$ and the last-mentioned product is positive,
then $S^2$ and $T^2$ supersymmetries are exact, while the other
two are broken. If $m^1  \Lambda^3_1$ is negative, then
$S^1$ and $T^1$ supersymmetries are exact provided the energy of the
ground state is equal to  $- m^1  \Lambda^3_1$.

{\bf One supersymmetry exact}
The case of the three -- quarters breaking of supersymmetry
is possible if the dimension of $N=4$ SUSY QM
is at least two ($D \geq 2$).
Indeed, for $D=2$ let us keep  the following set of parameters nonvanished:
$\Lambda^3_1, \Lambda^1_2, m^1$  \quad $Re (n^2) \equiv N^2$ .
After the further choice 
$m^1  \Lambda^3_1 =  \Lambda^1_2 N^2$
 one obtains
$$
\{ S^1 , S^1 \} = \{ S^2 , S^2 \} = 2H
$$
\begin{equation}
\{ T^1 , T^1 \}=2H + 4 m^1  \Lambda^3_1  \quad
\{ T^2 , T^2 \}=2H - 4 m^1  \Lambda^3_1 
\end{equation}
Therefore only $T^2$ supersymmetry is exact, while
all others are broken if the energy  of the ground state is
equal to $2m^1  \Lambda^3_1$,  and $m^1  \Lambda^3_1 > 0$.
 If $m^1  \Lambda^3_1$ is negative, then  $T^1$ is exact
provided the energy of the ground state is equal to
$- m^1  \Lambda^3_1$.

{\bf Three supersymmetries exact}
The situation of the one -- quarter
breaking of supersymmetry  can exist, if we add to the consideration
one more dimension, i.e., consider the
three -- dimensional $D=3$ $N=4$ supersymmetric quantum mechanics.

  Keeping the following
set of the parameters nonvanished
$\Lambda^3_1, \Lambda^1_2, \Lambda^2_3, m^1,  N^2$   
and $Im (n^3) \equiv M^3$,
under the conditions
$m^1  \Lambda^3_1  =  \Lambda^1_2 N^2 = - \Lambda ^2_3 M^3$
and $m^1  \Lambda^3_1 < 0$
we have
$$
\{ S^1 , S^1 \} = \{ S^2 , S^2 \} = \{ T^1 , T^1 \} = 2H 
+ 2 m^1  \Lambda^3_1 
$$
\begin{equation}
\{ T^2 , T^2 \} =2H - 6 m^1  \Lambda^3_1,
\end{equation}
then $T^2$ supersymmetry is broken, while all others are exact
under the condition that the energy of the ground state is equal
to $- m^1  \Lambda^3_1$. If the last -- mentioned 
 product is positive, then
$T^2$ supersymmetry is exact, while all others are broken
provided that the energy of the ground state is
$3 m^1  \Lambda^3_1$ and we arrive at the three-dimensional
generalization of the case of three quarter breaking of supersymmetry.
It is  obvious that all these cases can be obtained from the
higher dimensional ($D \geq 3$) $N=4$ supersymmetric quantum mechanics.

To summarize  one should note that according to the given
general analysis of partial SUSY breaking in the $N=4$ multidimensional
SUSY QM, there exist possibilities of constructing the models with
$\frac{1}{4}$, $\frac{1}{2}$ and $\frac{3}{4}$ supersymmetries
unbroken, as well as models with totally broken or totally
unbroken supersymmetries. However, the answer to the question
which  of these possibilities can be realized for the considered
system, crucially depends  on the form of the chosen superpotential
and on the imposed boundary conditions of the quantum mechanical
problem. 

 The  problem left opened is finding of the class of superpotentials 
$A(\phi^i)$
which lead to N=4 supersymmetric
Calogero and Calogero -- Moser models \cite{WYL},
which are related to the RN black hole quantum mechanics and to
$D=2$ SYM theory \cite{GT}.
Another topic --
the application of the given technique of N=4
SUSY QM to the description of the particle dynamics on AdS metric
will be given in subsequent publication.

We would like to thank  E. A. Ivanov for the helpful and stimulating
discussions and A. V. Gladyshev and C. Sochichiu for some comments.
 Work of A. P. was supported in part by INTAS Grant 96-0538
and by the
Russian Foundation for Basic Research,
Grant 99-02-18417.
Work of M. T. was supported in part by INTAS Grant 96-0308.
J. J. R. would like to thank CONACyT for the  support under the
program: Estancias Posdoctorales en el Extranjero, and Bogoliubov
Laboratory of JINR for hospitality. \\

\end{document}